\documentclass[aip,rsi,reprint,graphicx]{revtex4-1} 

\usepackage{epsfig}
\usepackage{xspace}
\usepackage{color}%

\usepackage{amssymb}

\draft

\begin{document}

\title{Atom Interferometry in Space: Thermal Management and Magnetic Shielding}

 \author{Alexander Milke}
 \author{Andr\'{e} Kubelka-Lange}
 \author{Norman G\"urlebeck}
 \email{norman.guerlebeck@zarm.uni-bremen.de}
 \author{Benny Rievers}
 \author{Sven Herrmann}
 \affiliation{Center of Applied Space Technology and Microgravity
(ZARM), University Bremen, Am Fallturm, 28359 Bremen, Germany}
 \author{Thilo Schuldt}
 \affiliation{DLR Institute for Space Systems, Robert-Hooke-Str. 7,
			 28359 Bremen, Germany}
  \author{Claus Braxmaier}
\affiliation{Center of Applied Space Technology and Microgravity
(ZARM), University Bremen, Am Fallturm, 28359 Bremen, Germany}
 \affiliation{DLR Institute for Space Systems, Robert-Hooke-Str. 7,
			 28359 Bremen, Germany}


\date{\today}

\begin{abstract}
Atom interferometry is an exciting tool to probe fundamental physics. It is
considered especially apt to test the universality of free fall by using two
different sorts of atoms. The increasing sensitivity required for this kind of
experiment sets severe requirements on its environments, instrument control, and
systematic effects. This can partially be mitigated by going to space as was
proposed, for example, in the Spacetime Explorer and Quantum Equivalence Principle Space Test (STE-QUEST) mission. However, the requirements on the instrument are still very challenging. For example, the specifications of the STE-QUEST mission imply that the Feshbach coils of the atom interferometer are allowed to change their radius only by about $260\,$nm or $2.6\cdot 10^{-4}\,$\% due
to thermal expansion although they consume an average power of $22\,$W. Also
Earth's magnetic field has to be suppressed by a factor of $10^5$. We show in this
article that with the right design such thermal and magnetic requirements can
indeed be met and that these are not an impediment for the exciting physics
possible with atom interferometers in space.
\end{abstract}

\pacs{03.37.-b,07.20.-n,07.55.Nk,07.87.+v}

\maketitle 

\section{Introduction}

One of the biggest challenges in current theoretical physics is that of finding
a valid theory of quantum gravity. Although many theories were proposed,
ultimately this question has to be resolved experimentally. Thus, many
experiments regarding tests of the fundamental assumptions of gravity and its
basic effects were carried out and suggested including tests of the Universality
of Free Fall (UFF). This basic principle is a cornerstone of Einstein's theory
of gravity and implies that the trajectories of freely falling, structureless
test particles in a gravitational field only depend on their initial position
and velocity.
In particular, the path of such test particles is independent of their
composition. A violation of this principle measured by the E\"otv\"os ratio
would help to discriminate between the different proposed theories of quantum
gravity and it would point in the right direction for further theoretical and
experimental development. If the UFF holds to the tested accuracy, bounds will
be placed on the viable alternative theories.

A very thriving tool to test the UFF using quantum matter is dual species atom
interferometry,\cite{Peters_2001,Fray_2004,Bonnin_2013} where two atomic clouds
propagate freely and their trajectories are compared. Although the advances
along this avenue are tremendous, very stringent requirements have to be
satisfied to reach down to an accuracy of the UFF test comparable to the
classical tests like lunar laser ranging\cite{Williams2012CQG} and torsion
balance experiments.\cite{Schlamminger08PRL} In the present paper, we show that
the requirements regarding the magnetic field and the thermal requirements,
which are closely tied to each other, can indeed be satisfied. As a source for
specific requirements we refer to the Spacetime Explorer and Quantum Equivalence
Principle Space Test (STE-QUEST) mission, which is a medium-size candidate
mission in the Cosmic Vision Program of the European Space Agency.
Different aspects of the STE-QUEST mission are described in Ref.\
\onlinecite{Hechenblaikner_2013,Aguilera_2013,Tino_2013,Schuldt_2014}. The
specific requirements for STE-QUEST are derived in Ref.\
\onlinecite{Schubert_2013}.

The STE-QUEST atom interferometer was designed with heritage from projects such
as the DLR project QUANTUS\cite{Zoest_2010} (QUANTengase Unter
Schwere\-losig\-keit) and the CNES project I.C.E.\cite{Nyman_2006}
(Inter\-f\'erom\'etrie Coh\'erente pour l'Espace). In QUANTUS, experiments with
Rubidium (Rb) Bose-Einstein condensates are carried out in the drop tower at ZARM in Bremen
and are currently prepared for a sounding rocket by the end of 2014.\cite{MAIUS}
A dual species interferometer is already included in I.C.E., where the
experiment is performed in parabolic zero-g flight.

The experimental sequence foreseen for STE-QUEST, is to cool two ensembles of
\textsuperscript{85}Rb and \textsuperscript{87}Rb atoms down to Bose-Einstein
condensation in two steps. First atoms are loaded from a magneto-optical trap
into a magnetic trap on an atom chip for evaporative pre-cooling. Then, they are
loaded into an optical dipole trap for final evaporative cooling. During this
step a strong homogeneous magnetic field of $160\,$G is applied in order to tune
atomic interactions using a so called Feshbach resonance. 
The magnetic field is generated by two coils in Helmholtz configuration with a middle distance of $105\,$mm, a coil's mean diameter of $206\,$mm, and equal
electrical current of $4.8\,$A flowing in the same direction. 
This way, two Bose-Einstein condensates of $10^6$ atoms shall be generated and released from
the trap to fall freely in Earth's gravitational field. A detailed description
of the atom interferometer design of STE-QUEST can be found in Ref.\ 
\onlinecite{Schuldt_2014}.

The article is organized as follows: In Sec.\ \ref{sec:requirements}, the
thermal requirements and the requirements for the magnetic shielding are summarized for
the example of the STE-QUEST atom interferometer. Subsequently, the methods
and models used for the analysis of the thermal control system and the magnetic
shielding are discussed in detail. The results are summarized in Sec.\
\ref{sec:results}.

\section{Thermal and Magnetic Shielding Requirements}\label{sec:requirements}

The sensitivity of a test of the UFF using dual species atom interferometry depends 
on how well spurious differential accelerations can be suppressed below the shot noise level of 
$2.9 \cdot 10^{-12}\,$ m/$\text{s}^{2}$ \ \cite{Schubert_2013}. In STE-QUEST, one such differential acceleration 
signal comes from the fact that both atom species couple differently to a magnetic field via the linear and
quadratic Zeeman effect. During the interferometer sequence a homogeneous magnetic quantization field of $B_{0} = 100\,$nT 
is applied and spatial variations of this field over the interferometer baseline of $12\,$cm have to be kept below $1\,$nT \cite{Schubert_2013}. Additionaly, an even stronger requirement of 
$\delta B/dz < 0.3\,$ nT/m has to be met during the prepartion of the Bose-Einstein condensates, where a strong homogeneous Feshbach field is required.  
The source of such magnetic field variations can be an external one like Earth’s magnetic field. 
As a consequence, the experiment has to be shielded against its influence. On the other hand, magnetic field
fluctuations and gradients produced by the instrument itself during the preparation of the Bose-Einstein condensates 
appear inside the magnetic shielding. These depend on the precise geometry, which can be affected by
thermal expansion, and an elaborate thermal control system has to be designed and verified to keep this systematic effects below the required
levels \cite{Schubert_2013}. We discuss both issues, the thermal control as well as the magnetic shielding, subsequently.

\subsection{Thermal Control System}
\label{thermrequ}
The most critical issues for the sensitivity of an atom interferometer
originating from thermal effects are temporal and spatial fluctuations of the
magnetic fields, which are used to manipulate the atomic ensemble. These
fluctuations are mainly caused by changes in position and length of the magnetic
coils, in particular, the Feshbach coils resulting from thermal expansion. The
thermal stability requirements have to be guaranteed during the entire
measurement cycle.  
To reach the aforementioned requirement for the magnetic field gradient of $\delta B/dz < 0.3\,$nT/m, 
the Feshbach coils have to be designed such that they are very close to Helmholtz configuration. To achieve this, 
a numerical simulation based on the Biot-Savart law was performed with 400 windings for each coil. 
The packing of the windings is essential in these simulations. 
Starting from an optimal position of distance and coil diameter for all windings found numerically, 
the effects of thermal expansion are tested along the z-axis in an area with a diameter of 
$40\,\mu$m placed symmetrically between the coils. The BEC will be located in that region during preparation, 
when the Feshbach coils are used. 
The temperature stability requirement is calculated due to maximum allowed position change from the ideal Helmholtz 
configuration of the coil's middle distance ($105\,$mm) and middle diameter ($206\,$mm) by linear thermal expansion 
of the copper windings, Torlon mounting pads, aluminum holdings and the titanium vacuum chamber (see Tab.\ref{tab:thermrequire}).

\begin{table}[h!]
\begin{center}
\begin{tabular}{p{1.8cm} p{2.6cm}p{2.3cm}}
 \hline
Component & & Requirement \\
\hline
FC & Diameter change & $\Delta d < 520\,$nm \\
FC & Distance change & $\Delta l < 320\,$nm \\
FC & Temp. stability & $\Delta T < 0.15\,$K \\
VC & Temp. stability & $\Delta T < 0.30\,$K \\
\end{tabular}
\caption{Requirements on the TCS for Feshbach coils (FC) and vacuum chamber
(VC) on which the Feshbach coils are mounted.}
\label{tab:thermrequire}
\end{center}
\end{table}

\subsection{Magnetic Shielding}

Fluctuating external magnetic fields can be a major source of systematic errors
in atom interferometry measurements. Also, small spatial inhomogeneities and
gradients of the magnetic field can lead to significant systematic phase shifts as explained in the beginning of Sec.\ \ref{sec:requirements}.
Thus, a high performance magnetic shielding is required for the targeted
performance of atom interferometers including the dual atom interferometer
foreseen in STE-QUEST.

The temporal fluctuations of external magnetic fields are mainly due to the
satellite's movement through Earth's magnetic field, $B_{Earth} \approx 50\,
\mu$T.  These fluctuations need to be shielded by at least a factor of
$S>10000$, i.e. to the low nT level. This shielding effectiveness
factor $S_{x,y,z}$ is defined by
\begin{equation}
S_{x,y,z}=\frac{B^{outside}_{x,y,z}}{B^{inside}_{x,y,z}}
\label{eq:S-Factor}
\end{equation}
with $B^{outside}_{x,y,z}$ as the outer magnetic field and $B^{inside}_{x,y,z}$
as the residual inner magnetic field inside the shielding.\cite{BurtPaper}

Even more critical than the shielding of temporal fluctuations is the
requirement on the spatial gradients of the magnetic field along the
interferometer baseline of $L = 12.4\,$cm. Two systematic effects need to be
considered here: a small differential acceleration of the two clouds after
preparation due to the linear Zeeman effect and the quadratic Zeeman shift
during the interferometer sequence. While mitigation strategies to alleviate
both effects have been identified in STE-QUEST,\cite{Schubert_2013} the
requirement on the magnetic field gradients to be met still remains challenging and no gradient
larger than $1\,$nT$/12\,$cm is allowed (larger than $1\,$nT over the baseline $L$).

The high shielding factor is reached by using a $\mu$-metal enclosure. This
magnetically soft nickel-iron alloy has a high permeability $\mu_r$ from $30000$
to $50000$, which results in a good shielding performance.
Shield designs with multi-layered $\mu$-metal shells can further improve the
performance of the shielding.\cite{BurtPaper} In addition to a passive
shielding with $\mu$-metal, an active compensation by coils can improve the
homogeneity and stability of the magnetic fields inside the
experiment.\cite{PharaoPaper}

\section{Methods and Modeling}

\subsection{Thermal Management}

The thermal management of an atom interferometer in space is challenging, since
the environmental condition may change rapidly due to orbital shadowing followed
by direct sunlight. Furthermore, convective heat transfer is negligible and heat
exchange within the system and with the environment has to be managed only by
radiation and conduction. In order to ensure the functionality of the atom
interferometer and its thermal control system (TCS) over the entire mission
period, the TCS has to be tested extensively in advance.

Further difficulties arise from the fact that the instrument itself produces high amounts
of waste heat. For example, the electronics and in particular the magnetic
coils are considerable sources of heat within the instrument. Thus, the TCS has to
be adjusted such that this heat is transported to the satellite's heat sink in
accordance with the requirements summarized in Sec. \ref{thermrequ}. A major
challenge for this is the cyclical instrument operation resulting in a
fluctuation of the magnitude of the heat fluxes. To ensure robustness and to avoid
disturbances on the instruments by pumps, which would yield artificial
accelerations beyond the acceptable level,\cite{Schubert_2013} the TCS has to
be designed as a completely passive subsystem.

In order to develop a design for the atom interferometer's TCS, which fulfills
the requirements listed in Tab.\ \ref{tab:thermrequire}, a 3D finite-element
model (FE) of the atom interferometer physics package has been set up in ANSYS
classic.\cite{ANSYS} This model is used for the definition of the thermal
interfaces between physics package and bus, for testing and validation of the
thermal stability of the design within operational requirements as well as for a
general optimization of the design with respect to perturbative influence of
thermal effects on the atom interferometer performance. The general approach
for the design of the TCS as well as details on the thermal aspects of the
instrument will be discussed in Secs. III A 1-2.

\subsubsection{General Approach}

The main task of the TCS of the atom interferometer is to assure a stable
operation of the instruments within the specified temperature requirements. Furthermore, the waste heat
generated by the instruments has to be transported to the satellite bus heat sink and the influence of
changing environmental conditions has to be buffered out.

In order to realize this, different options for thermal heat transport are
available. For the design of the TCS we considered the buffering of thermal
perturbations by means of phase change materials (PCM) as well as the transport
of heat by means of heat pipes and thermal straps. While PCM realizes a general
low perturbation level leading to a largely uniform temperature distribution,
the changing mass distribution during phase change introduces spurious
accelerations of the Bose-Einstein condensates due to changing gravity
gradients. A mitigation strategy for this would be possible by applying a filler
matrix. However, the complexity of the involved hardware would be considerably
increased. The use of heat pipes would provide a very efficient way to realize
high heat flows between instruments and heat sink. But it has to be considered
that heat pipes show a start-up phase and are working less efficient in
direction to the gravitational field. In addition to this, they show
self-gravity effects similar to PCM arising from the phase change of the
contained fluid during heat transport. Apart from this, the layout of a heat
pipe system, in particular the required diameters for the capillary and heat
pipe circle, would be challenging with respect to the requirements for magnetic
shielding of the atom interferometer. With these arguments in mind, heat strap
connections can be identified as the best option for thermal transport within
the atom interferometer and to the satellite bus. They transfer a considerable
amount of heat while offering flexible layout and dimensions within the
specified design requirements. Comparing to copper ($\lambda=360-400\,$W/Km) and 
nickel silver ($\lambda=26-40\,$W/Km), the carbon fibers show a much higher thermal 
conductivity ($\lambda=800\,$W/Km). Additionally the mass is reduced due to the very 
low density of those fibers\cite{techapp}.  

\begin{figure}[h!]
\begin{center}
\includegraphics*[width=8cm,angle=0]{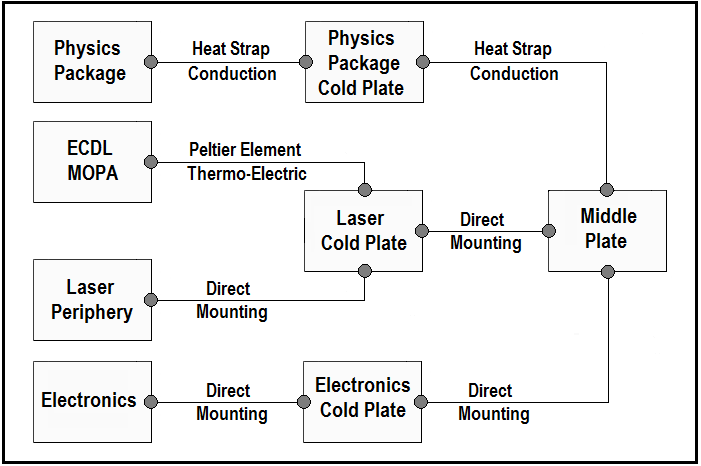}
\end{center}
\caption{General thermal control system approach for the atom interferometer.}
\label{figure1}
\end{figure}

A suitable design is shown in Fig.\ \ref{figure1}.
Because of the position in the bus and to avoid thermal couplings, the atom
interferometer is divided into four subsystems: the physics package, the
Extended Cavity Diode Laser (ECDL) and Master Oscillator Power Amplifier (MOPA),
the laser periphery and the electronic box. These subsystems are temperature
stabilized by means of three cold plates. The physics package's cold plate is
thermally connected with the middle plate, which is the heat sink of the
satellite, by a high conductive carbon fiber heat strap
through the $\mu$-metal. The middle plate is temperature stabilized to $(13 \pm
3) \,^{\circ}\mathrm{C}$ by the TCS of the satellite bus. The laser system and
the electronic box cold plates are mounted directly on the middle plate. Due to
the very strict temperature requirements of the ECDL and MOPA, they have to be
cooled actively by thermo-electric coolers.

The temperature stabilization of the laser system and the electronic box is a
well elaborated technique, which is realized in many laboratories
worldwide and will not be discussed here. The main focus is the thermal
management of the physics package and its instruments, which will be presented
in Sec. III A 2.

\subsubsection{Physics Package}

The TCS of the physics package is challenging especially concerning the magnetic
coils positioned inside the $\mu$-metal shield. On the one hand, the acceptable
thermal gradient over the coils is limited, since any thermal expansion affects
the magnetic field and, thus, the atom interferometer performance. Therefore,
any heat produced by the coil operation has to be transported quickly out of the
system. The coils produce a large amount of heat demanding a high heat transport
rate, which can be realized by increasing the size of the thermal interfaces. On
the other hand, a large diameter of the feed-throughs decreases the
shielding effectiveness factor $S$ of the $\mu$-metal. Hence, the maximum size
of the heat strap connecting the physics package and the middle plate is limited.
Nevertheless, the discussion in Sec.\ \ref{shield_fact} points out that a
feed-through diameter size of $40\,$mm for a thermal connection is acceptable
with respect to magnetic shielding in our design and keeps thermal deformation
within the acceptable range.
Furthermore, the main heat sources are not operating constantly, but periodically at
specific times during a $20\,$s measurement cycle. The main heat sources
included in the model and their individual operation time are listed in Tab.
\ref{tab:cycle}. This time-varying schedule is especially demanding concerning
the temperature stability requirement of the instruments and the mitigation of
the deformation of the Feshbach coils by thermal expansion. In order to keep
the temperatures of the physics package components stable, a set of high
conductive carbon fiber heat straps with individually designed diameter sizes
are applied for heat transport. Each of them is thermally connected to the
physics package cold plate, which is a temperature stabilized heat storage,
heat exchange and thermal interface mounting unit.

Due to a limited influence on the overall heat budget, thermal radiation from
the instrument to the magnetic shield and thermal conduction through the holding
structure is neglected in this model allowing for a higher computational
performance. Since these additional minor heat transport mechanisms would only
increase the overall heat exchange and homogenize the system thermally, they
would even improve the performance of the TCS. In this sense, the model
implements a thermal worst case scenario.

\begin{table}[h!]
\begin{center}
\begin{tabular}{p{2.1cm}p{1.6cm}p{1.8cm}p{1.75cm}p{1.75cm}}
\hline
Component & Operation  Time [s] & Peak Power [W] & Averaged Power [W] \\ \hline
Meso U & 0-2 & 22.5 & 2.25 \\
2DMOT & 0-2 & 13 & 1.3 \\
Offset 1 & 0-2 & 2x32.5 & 2x3.25 \\
Meso H & 2-2.1 & 50 & 0.25 \\
Base-chip & 2-2.1 & 50 & 0.25 \\
Science-chip & 2-5.6 & 8 & 1.44 \\
FCs & 5.6-8.9 & 2x67.5 & 2x11.1375 \\
CCD Camera & 10-20 & 5 & 2.5 \\
Dispenser Heater & 0-20 & constant on $ 40\,^{\circ}\mathrm{C} $ & constant on
$ 40\,^{\circ}\mathrm{C} $
\end{tabular}
\caption{Measurement cycle of $20\,$s of the atom interferometer and heat
loads generated by the instrument's subsystems, which are included in the
FE-model.}
\label{tab:cycle}
\end{center}
\end{table}

\begin{figure}[h!]
\begin{center}
\includegraphics*[width=8cm,angle=0]{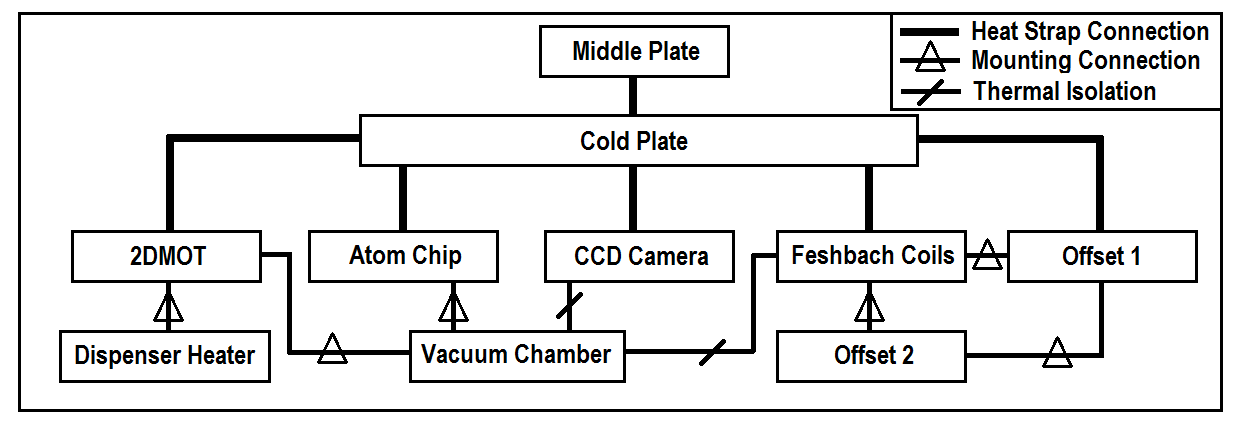}
\end{center}
\caption{Thermal control system design of the physics package.}
\label{fig:thermconcept}
\end{figure}

The thermal design of the physics package is shown in Fig.\
\ref{fig:thermconcept}. In this design, only the main heat sources and high mass
components, which are dominant for the overall heat budget are considered.
The thermal connections are shown schematically as heat strap connection (thick
line), mounting connection (triangle line) and thermal insulation (crossed out
line). One of the critical parts of the physics package is the vacuum chamber
with attached telescopes for interferometry, cooling, and detection. As
mentioned in Sec.\ \ref{thermrequ}, a thermal expansion of the vacuum chamber
has to be minimized. In order to satisfy this requirement, the vacuum chamber is
thermally isolated by Torlon pads\cite{Torlon} to the Feshbach coils as well as
to the CCD camera telescope. The highly conductive carbon fiber heat straps are
connecting the main heat sources to the physics package cold plate. The diameter
of each heat strap has been individually calculated. The heat transport through
the mounts and screws between the different components is considered here by
means of idealized FE models as well to assess the thermal properties of the
individual mechanic interfaces.

\begin{figure}[h!]
\begin{center}
\includegraphics*[width=8cm,angle=0]{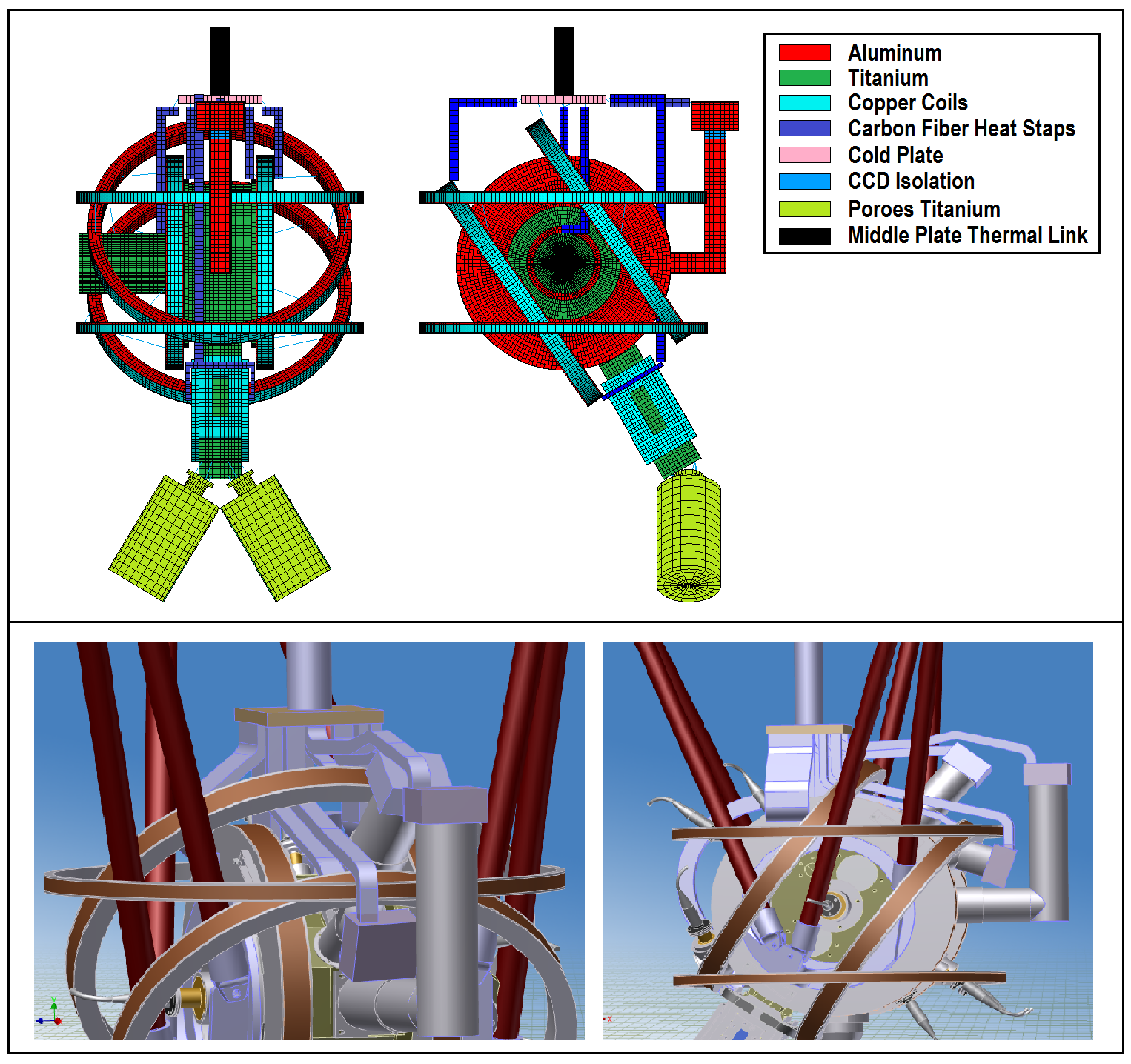}
\end{center}
\caption{\textbf{Top}: Meshed FE-model of the atom interferometer physics package.
\textbf{Bottom}: CAD-model of the atom interferometer physics package with the thermal interfaces.}
\label{fig:FEmodel}
\end{figure}

Based on this design and the geometries of the atom interferometer physics package, a
FE-model has been created (see Fig.\ \ref{fig:FEmodel} top). This
FE-model is an idealization of the physics package geometry design. The complex
geometrical structure of the telescopes have been idealized by representative
bodies implementing equivalent geometric diameters and component masses as
defined by the STE-QUEST study design.\cite{Schuldt_2014}
Furthermore, we take the influence of the wire isolation material and the gaps
between the wires into account by assuming that the magnetic coils (light blue)
are made of copper with reduced conductivity ($\lambda = 80\,\mathrm{W/mK}$) and
reduced density ($\rho = 5800\,$kg/m$^{3}$). The dispenser heaters are assumed
to be of porous titanium with a reduced conductivity ($\lambda = 1.5\,$W/mK) and
a reduced density ($\rho = 1526\,$kg/m$^{3}$). The vacuum chamber is made of
titanium (green) and the coil holders are made of aluminum (red). The coil
holders are used for mounting the heat straps and as a thermal interface of the
coils. The heat straps (ultramarine) are of highly conductive ($\lambda =
800\,$W/mK) \cite{techapp} and flexible carbon fibers as well as the connection to the middle
plate. The cold plate (pink) is of highly conductive copper ($\lambda =
390\,$W/mK). The atom chip, which is inside the vacuum chamber is connected with
the TCS via a copper gasket on the vacuum chamber. The insulation between the
Feshbach coils and the vacuum chamber is realized by Torlon pads. The charge-coupled device (CCD) camera
is mounted inside an aluminum box on top of the telescope, which is thermally
isolated by a layer of Torlon (dark blue). The middle plate is temperature
stabilized to $(13 \pm 3) \,^{\circ}\mathrm{C}$ by the TCS of the satellite bus.
The mounting connections between the components are represented by direct
node-to-node 3D-LINK elements (blue lines). The whole system has to be
thermalized from a uniform temperature of $13\,^{\circ}\mathrm{C}$ first to
reach equilibrium temperatures of the components and to achieve the necessary
temperature gradients for the heat fluxes. By means of an optimization of the
thermal performance of this FE-model the required individual heat strap's
diameter size is calculated and included in the computer aided design (CAD) model (see Fig.\
\ref{fig:FEmodel} bottom).

\subsection{Magnetic Shielding}
\begin{figure}[h!]
\begin{center}
\includegraphics[width=8cm,angle=0]{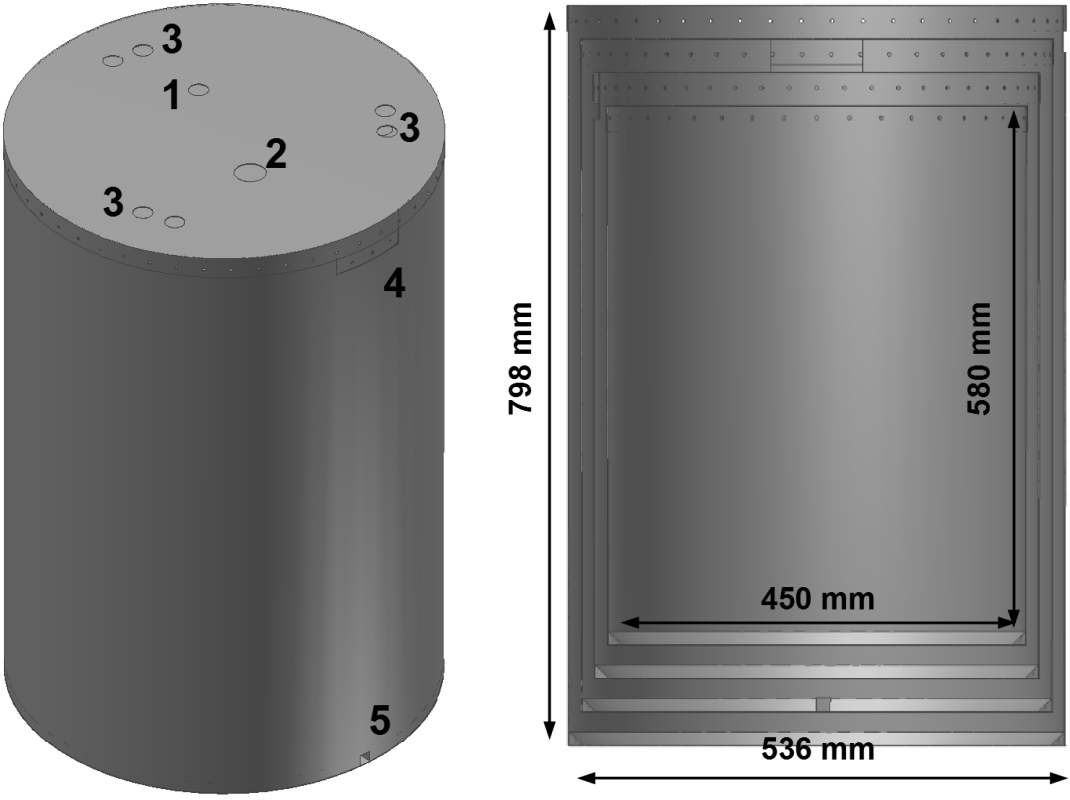}
\end{center}
\caption{\textbf{Left}: View of the complete magnetic shielding with
visible feed-through openings: 1) vacuum pipe, 2) heat strap, 3) mounting rack,
4) cables and fibers, 5) degaussing coil. \textbf{Right}: Cross-section of the
complete 4-layer magnetic shielding (necessary spacers not shown).}
\label{fig:Shielding}
\end{figure}

Based on previous FE simulations for the MAIUS and PRIMUS
projects\cite{MAIUS,PRIMUS} a shielding design with four individual layers of
$\mu$-metal was designed in CAD. In this model, each layer has a thickness of
$1\,$mm and consists of two parts, a cylindrical shell with welded bottom plate
and a cover with welded overlap (as shown in Fig.\ \ref{fig:Shielding}). The
best shielding factor was achieved in simulation for gaps between the
$\mu$-metal layers of $13\,$mm in radial and $35\,$mm in axial direction while
at the same time keeping the net mass of the magnetic shield as small as
possible. This shielding design provides several feed-throughs for the TCS and
supplies of the atom interferometer. An opening for a vacuum tube ($25\,$mm
diameter), which connects the pump section with the vacuum chamber inside the
magnetic shield, is foreseen on the top of the shielding. Next to this
feed-through, we consider an opening ($40\,$mm diameter) for the heat straps,
which connect the experiment to the middle plate, cf. Sec.\ \ref{thermrequ}.
Six more feed-throughs ($33\,$mm diameter) are implemented in the design on the
top of the shields to account for the mounting structure, which connects the
instrument to the middle plate. All openings are drilled through all four
layers. On the side of the cylindrical $\mu-$metal layers, a rectangular
feed-through ($100\times55\,$mm$^2$) is implemented for electrical cables and
optical fibers. The rectangular openings are rotated by $90^\circ$ to each other
to compensate magnetic leakage through the opening inside the shielding.
Additionally, an opening for a demagnetization cable is provided on the side in
the lower region of the cylindrical layers. The overall mass of the complete
shielding including mounting material is approximately $53\,$kg.

The shielding properties of this design were analyzed by means of FE
simulation, too. To simulate the effect of the Earth's magnetic field, the CAD
model of the shielding was placed inside a virtual enveloping body, which generated a
magnetic field of $B=40\,\mu$T. The residual magnetic field inside the magnetic
shielding is evaluated within virtual small test bodies at the position of the
atom interferometer in the middle of the shielding. The permeability of the
$\mu$-metal was set to $\mu_r=30000$ (manufacturer information\cite{Sekels}).

\section{Results}
\label{sec:results}

\subsection{Thermal Control System}

\begin{figure}[h!]
\begin{center}
\includegraphics*[width=8.0cm,angle=0]{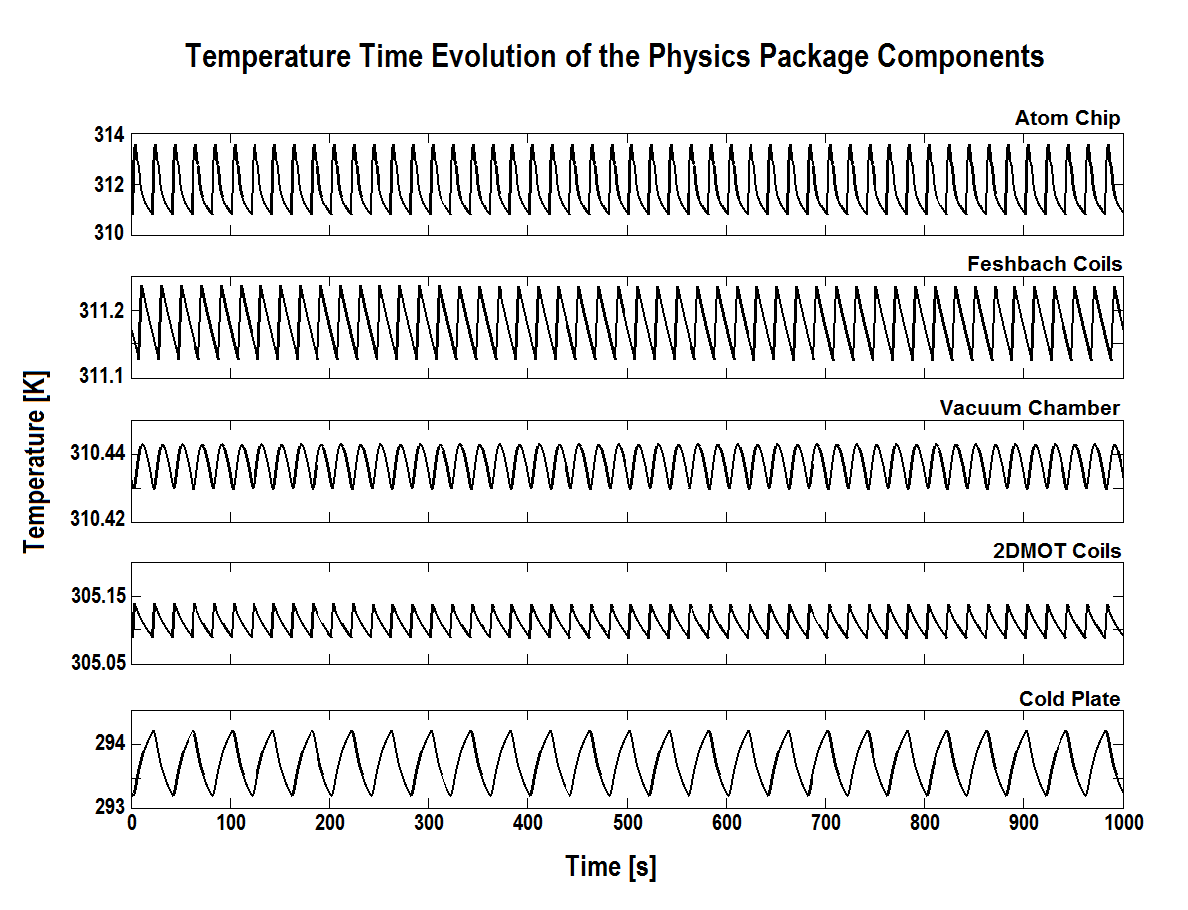}
\end{center}
\caption{Volume temperature time evolution of some critical components with
alternating environmental conditions $(\pm 3) \,^{\circ}\mathrm{C}$.}
\label{figure5}
\end{figure}

In order to test the TCS on changing environmental conditions, a variation of
the middle plate's temperature of $(\pm 3) \,^{\circ}\mathrm{C}$ within two
measurement cycles is considered. The temperature evolution of some critical
components is shown in Fig.\ \ref{figure5}. While the atom chip shows periodical
variations in the range of $2.5\,$K directly resulting from the chip operation, the most critical parts
like the Feshbach coils and the vacuum chamber show a rather stable temperature
evolution. The varying thermal conditions on the atom chip are demanding, since
thermal expansions arising from thermal gradients also lead to deformations of
the magnetic containment fields. While the in-plane expansion is uncritical
considering the isotropy and the low thermal expansion coefficient of the chip
substrate, the expansion of the copper mount below the chip may lead to a change
of the chip position during the measurement cycle. This effect can be mitigated
by a scaling of the chip currents but demands a thorough modeling and testing
with respect to the measurement cycle and the resulting thermal gradients.

Although the environmental temperature changes have an impact on the cold
plate's temperature, this effect is rather small on the other components.
In addition to the operational case $(13 \pm 3) \,^{\circ}\mathrm{C}$, a cold
case $(10 \pm 3) \,^{\circ}\mathrm{C}$ and a hot case $(16 \pm 3)
\,^{\circ}\mathrm{C}$ scenario is tested and the results are listed in terms of
thermal requirements in Tab.\ \ref{tab:thermresults}. The temperature shift in
the cold or hot case has no effect on the thermal stability
requirements. In summary, it can be stated that the TCS design presented and
tested by a FE-model is able to fulfill the thermal requirements, which are
set in Tab.\ \ref{tab:thermrequire}.

\begin{table}[h!]
\footnotesize
\begin{center}
\begin{tabular}{p{1.8cm}p{1.9cm}p{1.9cm}p{1.9cm}}
\hline
$\Delta T_{VC}$ in K & $\Delta T_{FC1}$ in K & $\Delta T_{FC2}$ in K\\
\hline
$0.014 \pm 0.004$ & $0.11 \pm 0.03$ & $0.11 \pm 0.03$\\
\end{tabular}
\caption{The maximal temperature variations of the vacuum chamber ($\Delta
T_{VC}$) and the two Feshbach coils ($\Delta T_{FC1}$ and $\Delta T_{FC2}$) for
a simulation time of $2000\,$s is shown. The result is the same for the operational, hot and cold case.}
\label{tab:thermresults}
\end{center}
\end{table}

\subsection{Magnetic Shielding}
\subsubsection{Shielding effectiveness factor}
\label{shield_fact}

The results of the FE analysis for the shielding effectiveness factors of the
discussed shielding design are summarized in Tab.\ \ref{tab:Shieldingfactors}.
It should be noted that in FE simulations of magnetic shieldings the modeled
shielding effectiveness factor is often better than in reality. This is caused
by multiple reasons such as variation within production tolerances, small gaps
between components or mechanical deformation of the assembled shielding. All of
this cannot be modeled accurately but strongly affect the actual shielding
effectiveness factor. We have, therefore, performed measurements on a prototype
magnetic shield of the above design but with three layers instead of four and
have shown an actual shielding effectiveness factor that is smaller by a factor
of four compared to the FE-modeling results. In order to account for this
discrepancy, the FE results for the STE-QUEST four layer shielding shown in
Tab.\ \ref{tab:Shieldingfactors} have been corrected by this factor. The
corrected numbers still indicate that the shielding effectiveness factor of
$10000$ along all spatial directions will be achieved.

\begin{table}[h!]
\begin{center}
\begin{tabular}{cc}
\hline
Component of $S$ & FE result\\
\hline
$S_x$ & $35000 \pm 1500$  \\
$S_y$ & $35200 \pm 1600$  \\
$S_z$ & $10700 \pm 500$
\end{tabular}
\caption{FE simulated shielding effectiveness factor $S$ for the three spatial
directions.}
\label{tab:Shieldingfactors}
\end{center}
\end{table}

\subsubsection{Magnetic Field Gradient}

\begin{figure}[h!]
\begin{center}
\includegraphics[width=8cm,angle=0]{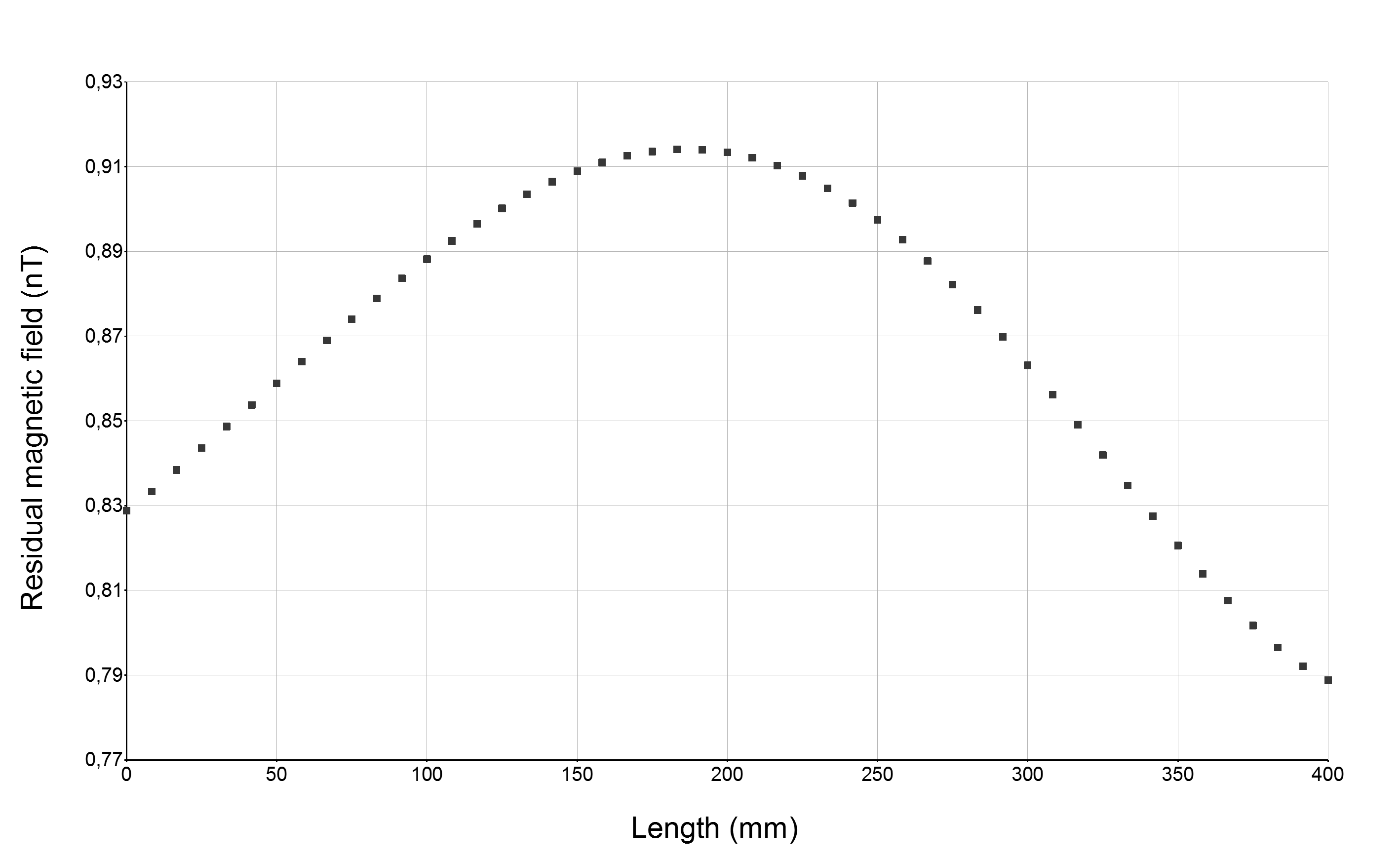}
\end{center}
\caption{Results of FE simulation of magnetic field gradients from an external
field $B^{outside}_{z}=40\,\mu$T along the cylinder axis. The gradient is
$<0.1\,$nT$/12\,$cm.}
\label{fig:Gradient}
\end{figure}

FE simulations along the cylindrical axis show that the gradient of the field
inside the shielding caused by outer magnetic fields is less than
$0.1\,$nT$/12\,$cm as shown in Fig.\ \ref{fig:Gradient}. This is a factor of
$10$ below the requirements given above. At this level, it can be assumed that field
inhomogeneities inside the magnetic shield will be dominated by residual
magnetization of the $\mu$-metal shells and not by residual external fields. It
is expected that with demagnetization of the magnetic shielding the effect of
this residual magnetization can be minimized to a gradient of less than
$1\,$nT$/12\,$cm.

\section{Conclusion}

While dual species atom interferometry in space is a powerful and promising tool
for fundamental physics science, the technological challenges that have to be
overcome in order to realize a sensor offering a high measurement accuracy are
considerable. Among these, the thermal management and the magnetic shielding of
the sensor hardware have a strong impact on the achievable sensor accuracy. In
this paper, we have analyzed the current design of the thermal control system
and the magnetic shielding with respect to the STE-QUEST mission specification
as a test case scenario. Both systems are designed and verified by means of
complex 3D FE models, which have been used to simulate the evolution of heat
flows, temperatures and magnetic field gradients for the dynamic boundary
condition throughout the STE-QUEST orbit and the measurement cycle. The numeric
results show that the current design of the atom interferometer allows to
satisfy thermal and magnetic requirements by a thorough design of both the
thermal control system and the magnetic shielding. Furthermore a more advanced design 
of the magnetic coils could help to relax the requirements on thermal expansion. 
The design and the FE models of both the thermal control system and the magnetic 
shielding will be used for future optimization or verification of overall design 
changes arising from other subsystem or mission requirements.

\begin{acknowledgments}
The authors acknowledge inspiring discussions with the members of the STE-QUEST
consortium in particular with Johannes Burkhardt and the team at Astrium GmbH in
Friedrichshafen.
This work was supported by the German space agency (Deutsches Zentrum f\"ur
Luft- und Raumfahrt -- DLR) with funds provided by the Federal Ministry of
Economics and Technology under grant numbers 50 OY 1302 and the European Space
Agency (ESA).
\end{acknowledgments}

\end{document}